\documentclass[10pt,conference]{IEEEtran}
\IEEEoverridecommandlockouts

\def\BibTeX{{\rm B\kern-.05em{\sc i\kern-.025em b}\kern-.08em
    T\kern-.1667em\lower.7ex\hbox{E}\kern-.125emX}}
\usepackage[utf8]{inputenc}
\usepackage{makecell}
\usepackage{amsmath,amssymb,amsfonts}
\usepackage{algorithmic}
\usepackage{graphicx}
\usepackage{textcomp}
\usepackage{xcolor}
\usepackage{listings}
\usepackage{tabularx}
\usepackage{array}
\usepackage{xspace}
\usepackage{url}
\usepackage{textcomp}
\usepackage{subcaption}
\usepackage{comment}

\definecolor{codegreen}{rgb}{0,0.6,0}
\definecolor{codegray}{rgb}{0.5,0.5,0.5}
\definecolor{codepurple}{rgb}{0.58,0,0.82}
\definecolor{backcolour}{rgb}{0.95,0.95,0.92}

\lstdefinestyle{mystyle}{
    backgroundcolor=\color{backcolour},   
    commentstyle=\color{codegreen},
    keywordstyle=\color{blue},
    numberstyle=\tiny\color{codegray},
    stringstyle=\color{codepurple},
    basicstyle=\ttfamily\footnotesize,
    breakatwhitespace=false,         
    breaklines=true,                 
    captionpos=b,                    
    keepspaces=true,                 
    numbers=left,                    
    numbersep=5pt,                  
    showspaces=false,                
    showstringspaces=false,
    showtabs=false,                  
    tabsize=2
}

\begin{document}

\title{\textbf{What do Computing Interns Discuss Online? An Empirical Analysis of Reddit Posts}}

\author{\IEEEauthorblockN{1\textsuperscript{st} Saheed Popoola}
\IEEEauthorblockA{\textit{School of Information Technology} \\
\textit{University of Cincinnati}\\
Cincinnati, USA \\
saheed.popoola@uc.edu}
\and
\IEEEauthorblockN{2\textsuperscript{nd} Ashwitha Vollem}
\IEEEauthorblockA{\textit{School of Information Technology} \\
\textit{University of Cincinnati}\\
Cincinnati, USA \\
vollemaa@mail.uc.edu}
\and
\IEEEauthorblockN{3\textsuperscript{rd} Kofi Nti}
\IEEEauthorblockA{\textit{School of Information Technology} \\
\textit{University of Cincinnati}\\
Cincinnati, USA \\
ntiik@ucmail.uc.edu}
}
\maketitle
\begin{abstract}

    Computing internships are the most common way for students to gain practical real-world experience. Internships have become a major part of most computing curricula because they help match student abilities or expectations with the demand of the workforce. The internship experience for students vary due to diverse factors such misconceptions about industrial realities, level of preparation, networking opportunities and so on. This papers attempts to provide insights into students discussions and opinions related to internship. We extracted 143,912 online Reddit posts related to computing internships out of 921,845 posts containing the rootword "intern", and then used topic modelling to unravel the common themes in the discussions. Next, we applied sentiment analysis techniques to understand the feelings expressed by the students in the Reddit posts. The results show that computing interns generally express a positive sentiment and the discussions were mostly related to academics, school admissions, professional career, entertainment activities, and social interactions.

\end{abstract}
\begin{IEEEkeywords}
interns, online forums, topic modelling, sentiment analysis
\end{IEEEkeywords}

\section{Introduction}

Traditional approaches to computing education often limit students' exposure and engagement to real-world projects; thereby, failing to fully harness their potential and creativity \cite{garousi2019aligning}, \cite{garousi2019closing}. The industry has often noted that fresh graduates have limited exposure to real-world projects and are often ill-equipped to deal with industrial projects \cite{brechner2003}, \cite{brng2013essence},\cite{ radermacher2013}, \cite{radermacher2014}. A major approach to exposing students to gain real world experience is through internships. 

 An internship is a program where students are trained in an industry to work on real-world projects \cite{kapoor2019}.The internship often occurs when the school is on break, typically in the summer.  A number of research has been conducted to understand the impact of internships on students \cite{martincic2009},\cite{minnes2021},\cite{kapoor2020}, factors that contibute to a (un)successful internship experience \cite{chang2016relationship},\cite{kapoor2020barriers}, and the challenges students faced during internships \cite{mia2020conceptual}. Majority of these approaches often used surveys or questionnaires which are limited in the number of participants and may not capture a wide range of students from diverse geographical locations.

 Online social media platforms such as Reddit provides an opportunity for users to share their thoughts on specific issues. A detailed analysis of how people behave in online forums may provide more insights into the fundamental mechanisms by which collective thinking emerges in a group of individuals. In 2022 alone, Reddit had 277 million posts with 1.43 billion comments \footnote{https://www.statista.com/topics/5672/reddit/\#editorsPicks}. Furthermore, Reddit data is relatively open compared to Twitter or Facebook \cite{baumgartner2020pushshift}. Thus huge amount of publicly available data  makes Reddit a popular a popular source of data for many research works \cite{medvedev2019anatomy},\cite{proferes2021studying}.

 This paper presents an analysis into the discussions of computing interns on Reddit platform. We extracted over 1400,000 posts related to computing internships and applied topic modelling techniques to gain insight into the common themes in the intern discussions. We then applied sentiment analysis on the post to extract the aggregate emotions in the posts. This paper makes the following major contributions.
 \begin{enumerate}
     \item A publicly available dataset \footnote{https://github.com/compedutech/Intern-Forum/tree/main/Reddit} on computing internships including over 143,912 posts (from June 2009 to December 2023) and the scripts used for data collection and data analysis
     \item A description of the common themes in computing internship discussion. This will provide insights into the major factors contributing to a (un)successful internship or the impact of the internship experience 
     \item A study on the sentiments expressed by interns in Reddit posts.
 \end{enumerate}
The contributions of this paper can lead to further research investigations into how to ensure students have a successful internship experience and satisfy the main objective of the internship program i.e, for students to gain practical real-world experience.

\section{Literature Review}
This section reviews the existing research on computing internships and data mining in computing education. These two fields provides the basis for empirical analysis of the online discussions of computing interns. This study relates to what computing interns discuss online by mainly focusing on the topics and sentiments expressed by the interns in Reddit posts.

Internships in general helps to bridge gap between theoretical knowledge students gain in formal academic settings and the real-world practical applications. Thus, internships enhances the career improvement, personal growth and intellectual curiosity of students \cite{tan2005incorporating},\cite{ordille2019internships}. The experience gained at internships also helps to build confidence and decent understanding of career choices \cite{mia2020conceptual},\cite{alkadi2008student}. 

Existing research have shown how internships help to gain practical knowledge, understand industry work environment and secure future employment opportunities \cite{kapoor2020barriers},\cite{jaime2019effect}. However, there is still limited knowledge about why some students chose not to intern before they work full time even though previous studies have shown that 40\% of students prefer to intern at least one time before they graduate \cite{kapoor2020}. Hence, there is need to understand student perception towards internships. Furthermore, internships have become an important part in job recruiting process as it allows companies to evaluate a student ability over a certain period \cite{kapoor2019}, \cite{summit2017positive}. 

Similarly, internships provide an opportuntiy for students to understand professionalism and workplace issues in the field of computing \cite{kubilus2000assessing}. The experience gained often help students to perform better in future. Binder et al. \cite{binder2015academic} conducted a study on about 15,000 students to examine the impact of internships on the students’ academic performance. Their results show that students who have completed the internships has better academic performance irrespective of student’s natural talent. This is consistent with the results by Jamie et al.\cite{jaime2019effect}. Another study also reveals that students who have done the internship for longer time have better academic performance compared to students who have done internship for shorter time \cite{blicblau2016role}.

Although computing internships benefit students in many ways, they also introduce various issues and challenges. For example, student must understand the work culture and act accordingly while applying their theoretical knowledge practically. To mitigate some of these challenges, Sweetser et al. came up with a design, implementation and evaluation of a computing internship which benefits both the employer and student \cite{sweetser2020setting}.

Data mining has become an important tool to analyze data, gain better understanding, and improve learning experience within the field of computing education \cite{algarni2016data}. The use of data mining techniques to analyze online discussions provides a better understanding of what students talk about and the challenges they face. Techniques such as sentiment analysis also helps to understand the student perception and their opinion on various programs, including computing internships \cite{shaik2023sentiment}. This provides deeper insights on how to improve these programs.

The hands-on experience students gain from internships will definitely help the students prepare better for their future career and enhances their professional growth. On the other hand the data mining techniques can help to analyze students data to improve learning \cite{barron2019emotion}. Therefore, by applying data mining techniques on internships data, researchers can have better understanding of interests of computing interns. These approaches help both the students and employers to create better internship programs. This papers advances the literature by analyzing interns discussion on Reddit platform via data mining techniques.

\section{Methodology}

This section discusses the research methodology adopted in this paper and the research questions we aim to answer.

\subsection{Data Collection}
We extracted Pushshift \cite{baumgartner2020pushshift} Reddit data dump \footnote{https://academictorrents.com/details/9c263fc85366c1ef8f5bb9da0203f4c8c8db75f4} from June 2009 to December 2023. We filtered the dataset to search for posts that contain the following four key words "intern","interns","internship", and "internships". We were able to extract 921,845 posts. However, this contain lots of noise that may not be related to computing internship. The following is an example of a post that contains the word "intern" but it is not related to computing internship

\textit{
Watching Bones for the first time and we're currently in Season 11.
(For the record, my non-intern favourites are Sweets, Caroline and, for all her annoying moments, Brennan; my least favourite is Cam. My favourite interns are Wendell and Clark, and my least favourites are Oliver Wells and Jessica Warren).}

The HuggingFace implementation\footnote{https://huggingface.co/facebook/bart-large} of the Facebook Bart model \cite{lewis2019bart} was used to remove unrelated posts by automatically classifying posts into computing or non-computing. We first evaluated the performance of the Bart model by randomly selecting over 100 posts from the database, then the authors independently classified the posts to obtain a ground truth. The authors used the following inclusion and criteria to determine if a post is related to computing or not. The inclusion criteria for computing post are.
\begin{itemize}
    \item Written by a computing major: This means if a person has a background in computing like mentioning computing major or related fields such as Information Technology,Computer Science,Computer Engineering,Software Engineering, etc. Persons having computing major most likely talks about topics related to computing.
    \item Company (or section of the company) in computing field: Posts form companies which are involved in computing related activities.
    \item Technical terms: Posts which contains technical terms or industry specific language related to computing field.
    \item Discussion related to computing technologies: Posts which discusses about any computing technologies, methodologies.
    \item Programming skills: posts which has code snippets or programming tutorials.
    \item References: posts which provides citations or mentioning about any articles, books, research papers related to computing topics.
\end{itemize}
The following exclusion criteria was used to classify posts as non computing.
\begin{itemize}
    \item At least one of the inclusion criteria for computing group is not met
    \item Experience in non-technical work: posts that share professional and personal experiences related to non-computing such as marketing, finance etc.
    \item Discussion related to non-computing fields: posts discussing about topics unrelated to computing fields.
    \item Discussion of non-technical challenges: posts discussing challenges or thing learning in areas outside the scope of computing like interpersonal relationships, travel experience. Basically, sharing the non-technical experiences
\end{itemize}
We only used the post where are all the classifications by the authors were exactly the same. The performance of the Bart model was then compared against the original classification by the authors. At the end of the initial evaluation, the model had a high accuracy in identifying non-computing related data as 90\% of the posts classified as non-computing were  correct. However, there were a lot of false positives in predicting the computing related posts as only 65\% of the posts classified as computing were actually computing related. At the end of the classification process, 308,841 posts were classified as computing and 613,004 posts were classified as non-computing. 

The Reddit platform is organized into forums called subreddits which are managed by independent moderators. Each subreddit often targets a specific theme and the subreddit moderators can set the
rules for joining, viewing, and posting to the subreddit. Furthermore, every post on Reddit is associated with exactly one subreddit, and the subreddit of the post indicates which subject the post belongs to. 

\begin{figure}
    \centering
    \includegraphics[width=0.9\linewidth]{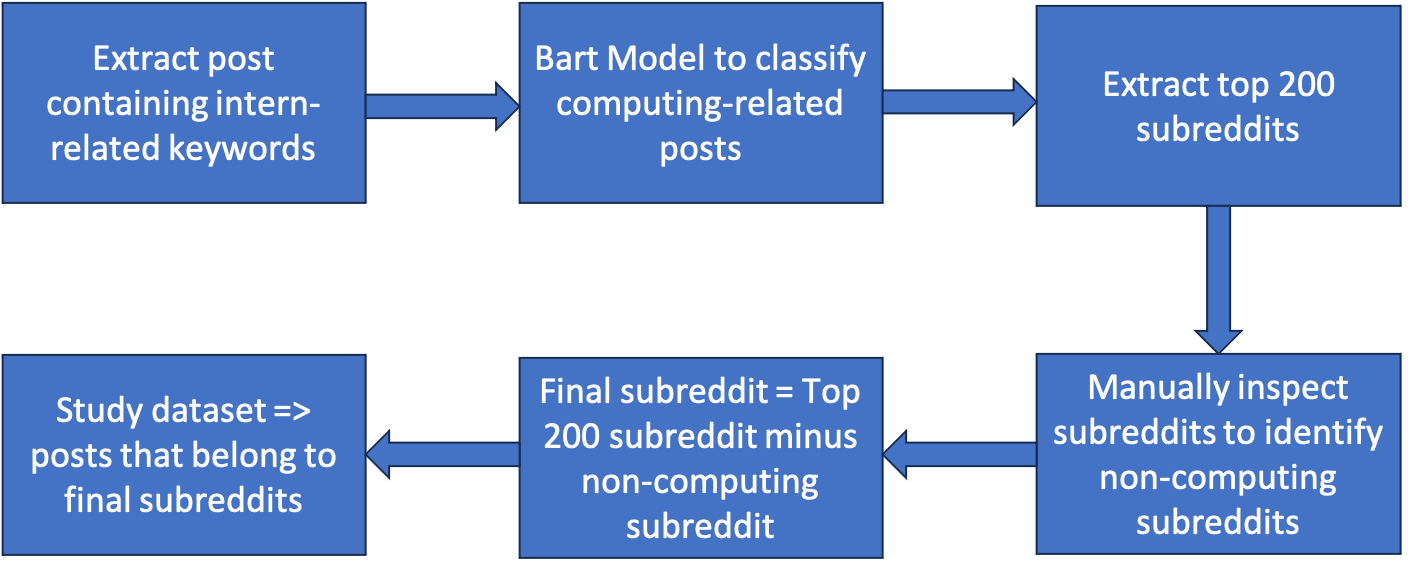}
    \caption{The Data Collection Process}
    \label{fig_data_collection}
\end{figure}

A high quality dataset is very important for getting correct results. Hence, to further enhance the quality of the dataset used for the study, we analyzed the subreddits for each of the 308,841 posts classified as computing related by the Bart model. We then extracts the top 200 subreddits with the highest number of posts. The top subreddits were related to CScareer, school admissions, relationships, specfic discipline (e.g., subreddits in software enginnering), and immigration. We then looked into the description of each subreddit and filter out subreddits related to relationships, subreddits reserved for mature audience, or subreddits dedicated to a non-computing discipline such as accounting. We chose to remove the posts related to relationships because we found out that majority of the posts on relationships were not related to computing. Furthermore, we did not remove generic subreddits like immigration and school admissions because a lot of computing interns also share their internship experience in these subreddits. At the end of the process, we extracted 143,912 posts out of the 308,841 posts originally classified as computing. We believe the combination of using Bart model for classification and filtering out unrelated subreddits produced a quality dataset that was used in this study. Figure \ref{fig_data_collection} provides a graphical description of the data collection process.

\begin{figure}
    \centering
    \includegraphics[width=1.0\linewidth]{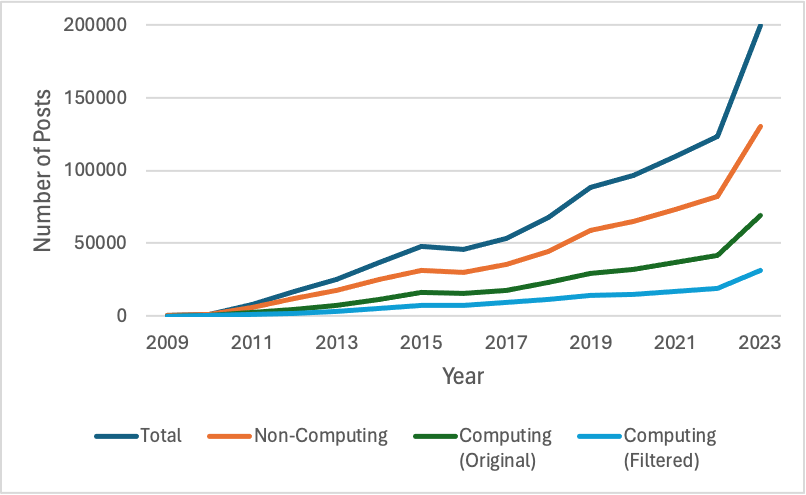}
    \caption{Number of Posts with Intern as a Root Word  }
    \label{fig_num_of_posts}
\end{figure}

\begin{figure}
    \centering
    \includegraphics[width=0.9\linewidth]{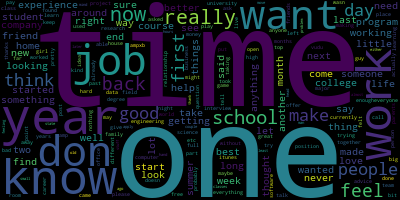}
    \caption{Word Cloud for unfiltered Computing-related Posts}
    \label{fig_word_cloud}
\end{figure}

\begin{figure}
    \centering
    \includegraphics[width=0.9\linewidth]{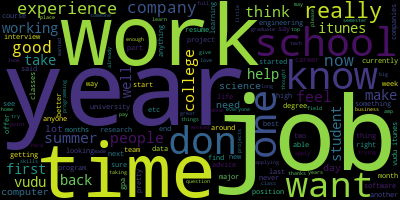}
    \caption{Word Cloud for Filtered Computing-related Posts}
    \label{fig_word_clean}
\end{figure}

Figure \ref{fig_num_of_posts} provides an overview of the total number of posts per year. It should be noted that the year 2009 only contain posts from June to December of 2009. The figure also shows that the number of internship related posts have been increasing every year, with a sharp rise around 2022 and 2023. Figure \ref{fig_word_cloud} contains the word cloud for the original computing-related post, while Figure \ref{fig_word_clean} contains the word cloud for the filtered posts. It can easily be seen that Figure \ref{fig_word_clean} contains much more relevant words, which may indicated that the filtered dataset is of a higher quality. It can be also seen from the Figure \ref{fig_word_clean} that Job, time of the year, and school are the prominent keywords. This is not surprising because the internships is a job experience that is usually done at a specific time of the year. For the word cloud, we removed the words "intern" or "internship" since every post is expected to contain these words.

The complete list of all the posts (including the original unclassified posts, unfiltered classified post, and filtered classified posts), the scripting codes used for filtering and classification, and the manually classified post by the authors are all provided in a public GitHub repository \footnote{https://github.com/compedutech/Intern-Forum/tree/main/Reddit}.

For the rest of this study, we focused on the 143,912 posts that has been filtered based on the subreddit. We believe this successfully capture many of the posts related to computing internships on Reddit.

\subsection{Research Questions}\label{RQ}
Online forums provide a platform for users to post questions and replies about different topics. A comprehensive analysis of the discussions related to a specific concept (e.g., internships) may provide insights into the patterns of thought and understand how collective thinking appears among a group of individuals. Hence, we may be able to know the core challenges interns face, what students may expect during an internship, and factors that contribute to a successful or unsuccessful internship experience.


The objective of this study is to gain an in-depth insight into online discussions of computing interns. Hence, the following three research questions guide the study.
\begin{itemize}
    \item \textbf{RQ1}: What are the prevalent themes in the discussions related to computing internship?
     \item \textbf{RQ2}: What sentiments are expressed in discussions related to computing internships?
     \item \textbf{RQ3}: What are the major domains in the discussions related to computing internships?
\end{itemize}

\subsection{Study Design}\label{design}
For  the study, we used topic modelling and sentiment analysis techniques to answer the research questions. The following subsections provide a detailed description of the study design.
\subsubsection{Overall Pipeline of Topic Modeling}
Topic modeling is a process that uses unsupervised machine-learning techniques to identify topics within a text without human input. This method is commonly used to uncover underlying thematic structures in the content of a text \cite{maier2021applying}. The process typically includes three main steps: pre-processing, topic modeling, and results interpretation.

\textit {Step 1: Data Pre-processing.} 
The first step is to refine the text documents via data preprocessing. Data preprocessing helps to reduce computational complexity and increase the accuracy to the topic modelling process. We followed these steps during data preprocessing phase. First, we converted the text to lower case and removed meaningless characters(such as '[' and '/'), punctuations, numbers, urls, and short words with one or two letters. Second, we used tokenization techniques to break down sentences into individual words and then remove common stop words (such as 'a','the','but', etc). Finally we used stemming  to reduce words to their base form (e.g., changing 'worked' to 'work').

\textit{Step 2:Topic Modeling.} The second step involves using machine learning algorithms to identify topics or themes within text documents. Common algorithms for this task include Latent Hierarchical Dirichlet Process (HDP) \cite{johnson1967hierarchical}, Semantic Analysis (LSA) \cite{dumais2004latent}, and Latent Dirichlet Allocation (LDA) \cite{blei2003latent}. For our analysis, we utilized the LDA implementation from the Gensim Python package \cite{rehurek2011gensim}. To determine the optimal number of topics, we compared the coherence scores for a range of topic numbers.

\textit{Step 3:Interpretation of Results}
The third step is to interpret the results by identifying the topics associated with the keywords. We used open-coding techniques on selected posts containing these keywords to name the identified topics. The ope-coding approach was conducted in two main steps. First, we extracted the top 20 keywords associated with each of the topic identified by the LDA algorithm. Secondly, we manually assign labels to the topics based on the top words. The labels were assigned after reading sample documents with high probabilities for a specific topic to understand the context better. The main goal of the labelling process is to summarize the prevailing themes for each topic.

\subsubsection{Sentiment Analysis}\label{sentiment}
Sentiment analysis, also known as opinion mining, is a field of natural language processing (NLP) that is used to determine the emotional tone in text \cite{oyebode2019social}. This analysis helps to understand the sentiment expressed in the text and they are usually classified into categories such as positive, negative, or neutral.

We used two popular Python packages TextBlob \cite{loria2018textblob} and Vader \cite{hutto2014vader}, to extract the sentiments in the computing related posts. Both Textblob and Vader will return an analysis score from -1 to 1, where -1 is extremely negative and 1 is extremely positive. We followed the guideline used by Oyebode and Orji \cite{oyebode2019social} to classify the posts based on the analysis scores. Table \ref{table1} show the rules for classifying the post based on the sentiment score returned by Textblob and Vader. The final classification rule adopted is that a post is classified as positive if both Textblob and Vader returned a positive classification, while the post is classified as negative if both Textblob and Vader returned a negative classification; otherwise, the post is classified as neutral.
\begin{table}\centering
    \begin{tabular}{|p{.35\columnwidth} |r |r|}
        \hline
        \multicolumn{1}{|c|}{Classification}  & \multicolumn{1}{c|}{Textblob} & \multicolumn{1}{c|}{Vader} \\
        \hline
        Positive & $>$ 0 & $>$ 0.5 \\
         \hline
        Negative & $<$ 0 & $<$ -0.5 \\
         \hline
        Neutral & 0 & $\ge -0.5$ \&\& $\le 0.5$ \\
         \hline
    \end{tabular}
    \caption{Classification Rules Based on Analysis Scores}
    \label{table1}
\end{table}

\section{Results}
This section presents the results of the data analysis discussed in the previous section. The results have been grouped based on the research questions introduced in Section \ref{RQ}. To answer these research questions, we have used topic modelling and sentiment analysis approaches described in Section \ref{design}.

\subsection{RQ1: What are the prevalent theme in the discussions related to computing internship}\label{rq1}
We used to coherence score \cite{syed2017full} of the LDA model to get the optimum number of topics by comparing the coherence scores of different LDA models from 2 to 20 topics. The results shows that intern discussions are centered around five main themes. Sample posts containing the keywords were extracted, read, and manually analyzed to better understand the context and interpret the topic modelling results. The analysis shows that intern discussions were centered around academics/admissions, professional career including job applications, entertainment activities, emotions/social interactions, and the core internship experience (mostly as software engineer). This results show that the interns view the internship experience as a vital part of the school admission or job application process. The interns personal life are also not dissociated from the internship experience, and they still had to relate socially with their colleagues. Finally, the results may also indicate that majority of the computing interns work in a software engineering role during their internship. Table \ref{table2} shows the keywords associated with each of these five topics.

\begin{table}\centering
    \begin{tabular}{|p{.28\columnwidth} |p{.63\columnwidth}|}
        \hline
        \multicolumn{1}{|c|}{Topics}  & \multicolumn{1}{c|}{Key words}  \\
        \hline
        Topic 1 -Entertainment & itunes, vudu, itunesport, movie, google, point, amp, war, play, season, star, anywhere, unrated, man, collect, day, split, playport, port, dark game \\ 
         \hline
        Topic 2 - Core internship experience & company work, experience, job, interview, get, offer, intern, engineer, data, year, like, apply, develop, software, position, learn, know, project,look 
        \\
         \hline
        Topic 3 - Social interactions & like, one, get, time, work, know, want, back, day, thing, people, could, look, make, said, think, try, got, ask, year
        \\
         \hline
        Topic 4 - Academics and admissions & 
        school, year, university, research, work, student, science, college, program, one, gpa, class, club, computer, intern, major, also, get, apply, math
        \\
         \hline
        Topic 5 - Professional career & get, job, work, year, like, time, want, feel, know, school, really, start, graduate, take, college, also, make, one, think, experience
        \\
         \hline
    \end{tabular}
    \caption{Topic and Keywords}
    \label{table2}
\end{table}

\subsection{RQ2: What sentiments are expressed in discussions related to computing internships}

\begin{figure*}[t!]
    \centering
    \begin{subfigure}[t]{0.32\textwidth}
        \centering
        \includegraphics[height=1.1in]{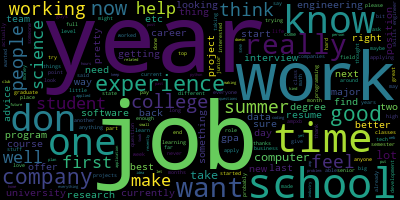}
        \caption{Positive Sentiment}
    \end{subfigure}%
    ~ 
    \begin{subfigure}[t]{0.32\textwidth}
        \centering
        \includegraphics[height=1.1in]{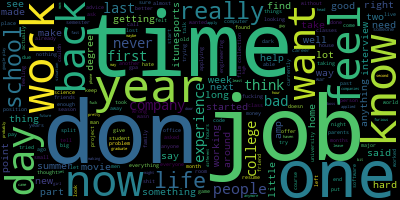}
        \caption{Negative Sentiment}
    \end{subfigure}
     ~ 
    \begin{subfigure}[t]{0.32\textwidth}
        \centering
        \includegraphics[height=1.1in]{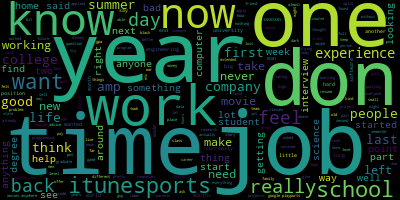}
        \caption{Neutral Sentiment}
    \end{subfigure}
    \caption{Word Cloud for Positive, Negative, and Neutral Sentiments}
    \label{sentiment_word_cloud}
\end{figure*}
Although the discussion topics in RQ1 is important for understanding prevailing themes, it does not capture how student feel about them. For example, some posts actually expressed frustration that their academic program required them to have an internship training but the students have not been able to secure one. From the context of the post, it appears that the frustration was directed towards the internship requirement and not their inability to secure an internship position. This shows that understanding the sentiments of the discussion is also very important.

The results of the sentiment analysis process described in Section \ref{sentiment} show that 104,634 (72.7\%) posts were classified as positive, 7,939 (5.5\%) posts were classified as negative, and 31,339 (21.8\%) posts were classified as neutral. This result indicates that majority of the interns were satisfied with their internship experience. Figure \ref{sentiment_word_cloud} shows the word cloud for each of the sentiments. The figure show that the words "year" , "job", and "work" are prominent for positive sentiments. This may indicate overall satisfaction with the time of the internship and the work they were assigned to do during the the internship period. The words "feel", "know" and "don (stemmatized don't)" were prominent with negative sentiment.  This may indicate the technical and emotional challenges the interns experience.

The overlap of job in both positive and negative sentiment also suggest that the internship experience for computing interns might be a mixed bag. Even though interns may value the opportunity and potential career benefits of internships such as landing future jobs (i.e., ‘job’ in positive sentiment), they might also encounter difficulties, anxieties, uncertainties, or dissatisfactions with workload or the actual work (i.e., ‘job’ in negative sentiment). This overlap highlights a potential area for improvement in internship programs. That is, internship programs can be designed to provide interns with valuable and engaging work experience that contributes to their growth while minimizing negative aspects like excessive workload or tedious tasks.

\subsection{RQ3: What are the major domains in the discussions related to computing internships}
 
The Reddit platform is made up of subreddits that targets a specific topic or domain. The description of each subreddit often specifies the expected domain for any posts created in the subreddit. Subreddits are typically created and moderated by users. Moderators enforce community-specific rules, manage content, and ensure discussions remain relevant, appropriate, and related to the specific subject area defined in the subreddit description. We analyzed the subreddits of the posts in order to understand the major subject areas that computing interns are interested in.

\begin{figure}
    \centering
    \includegraphics[width=1.0\linewidth]{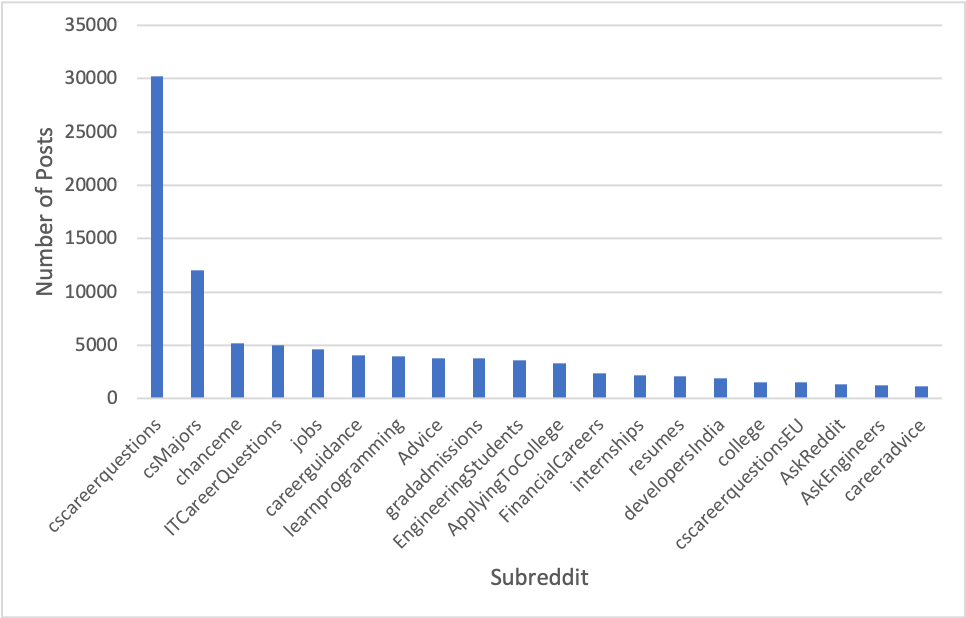}
    \caption{Number of Posts for Top 20 Subreddits  }
    \label{fig_subreddit_hist}
\end{figure}

The analysis of the subreddits for computing posts provided more insights into the intended subject matter of the posts. We extracted and manually analyzed all the subreddits associated with the collected data. The result show that there were 19,663 subreddits in the original unifiltered data and 159 subreddits in the filtered data used for this analysis. This shows that computing interns posts in a wide range of subject matter. The highest subreddit is "cscareerquestions" with 30,209 posts or about 20\% of the total post. According to the subreddit description,  cscareerquestions is \textit{for those who plan to enter or already in the computer science field to help them navigate the challenges of the industry and share strategies to be successful.} The next top subreddits are csMajors, chanceme (chances of acceptance to college), ITCareerQuestions, and jobs. Figure \ref{fig_subreddit_hist} shows the top 20 subreddits and the number of posts asscoiated with them. 

Next, we manually went through the description of each subreddit in order to extract the subject niche for that subreddit. We then identify subreddits in similar subject matters and manually analyze them to extract the major domains for the posts. The results of the analysis reveals eight major domains described below.
\begin{enumerate}
    \item Academics (including admissions). This domain captures all the subreddits related to academics including admissions, classwork, university life, and so on. Examples of subreddits in this domain includes gradadmissions, ApplytoCollege, CSEducation and university-specific subreddits such as Cornell, UTAustin, and Purdue. This domain has the highest number of subreddits.
    \item Career. This domain covers all the subreddits related to career including how to get jobs or internships, career pathways, job experience, building strong resume, work-life balance, and so on. Example of subreddits in this domain includes cscareerquestions,jobs, careerguidance,techjobs, etc. This domain has the highest number of posts.
    \item Data Science. This domain includes subreddit related to data science and machine learning. This doamin has the least number of subreddits and domain. Sample subreddits in this domain are learnmachinelearning, dataengineering, and datascience.
    \item Finance . This domain covers all subreddits related to financial advice or  financial companies. Examples of subreedits in this domain include FinancialCareers,Big4, and quant.
    \item General. This domain covers all other subreddits that are generic and not classified to a specific niche. Some of the subreddits are specific to specific countries. Examples of subreddits in this domain are Advice, AskReddit, LifeAdvice,India, and askSingapore.
    \item Programming and Software Engineering. This domain includes the subreddits that specifically targets programming, software and development, and web development. Although a significant number of software engineering-related posts are present in all other domain, we only include subreddits that targets only software engineering related activities in this domain. Examples of subreddits include learnprogramming, learnpython, typescript, and SoftwareEngineering
    \item Mental Health and Social Interactions. This domain includes subredits that focus on mental health and social interactions. Example of subreddits are mentalhealth, vent, socialanxiety, and therapists.
    \item Technical Competence. This subreddit covers all other technical subreddits (related to engineering or computer science) except for data science and software engineering. Examples of subreddits in this domain include sysadmin, InformationTechnology,techsupport, and AsksEngineers.  
\end{enumerate}

\begin{figure}
    \centering
    \includegraphics[width=1.0\linewidth]{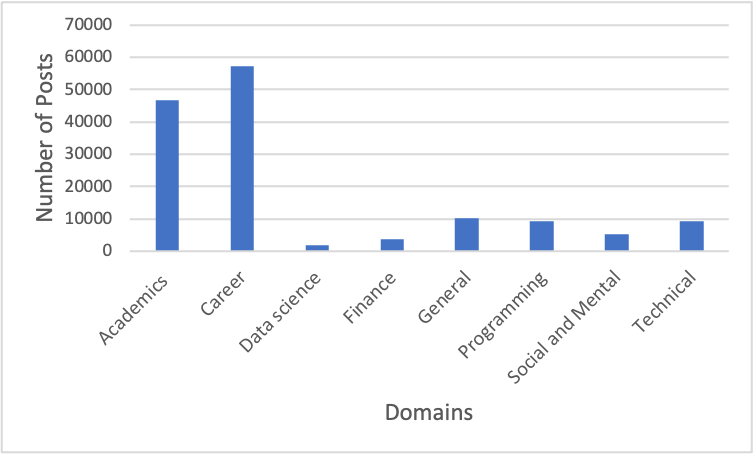}
    \caption{Number of Post in Each Domain  }
    \label{fig_subreddit_post}
\end{figure}
\begin{figure}
    \centering
    \includegraphics[width=1.0\linewidth]{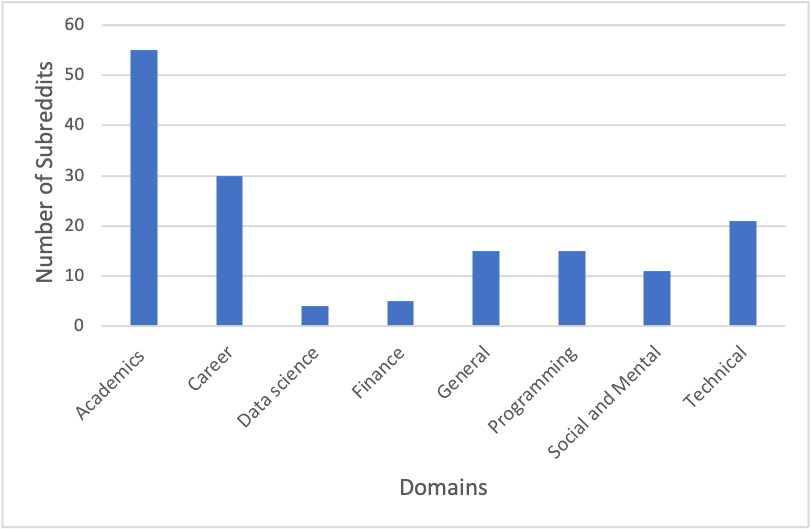}
    \caption{Number of Subreddits in Each Domain  }
    \label{fig_subreddit_domain}
\end{figure}

There are also a number of subreddits that belong to more than one domain. In this scenario, we group the subreddit to the more dominant domain. For example, developersIndia, a subreddit for software developers in India, is classified under the programming and software engineering domain instead of the general domain. Figure  \ref{fig_subreddit_post} and Figure \ref{fig_subreddit_domain} presents an overview of the number of posts and number of subreddits in each domain respectively. It can be seen that majority of the posts and subreddits in the domains are related to career, seeking technical knowledge, academics, and admissions. This is also consistent with the RQ1 results in Section \ref{rq1}.

\section{Implications for Computing Curriculum}
The results presented in this study reveals the specific concerns that computing interns often face. These results can be used to guide the software engineering curriculum  by including the following guidelines.
\begin{enumerate}
    \item Emphasis on career preparation and guidance.  The results (especially RQ3) show that students  often seek career advice such as how to navigate internship applications and transition to the industry. We believe that there is a need to improve the curriculum by including career counselling and professional development that covers resume building, interview preparation, techniques for navigating job offers and negotiation, and career paths within the computing fields such as industry, academia, research labs, or startups. The computing departments may also need to increase collaboration with the career services.
    \item Integration of soft Skills and real-world problem solving. Many of the discussion also centers around soft-skills like communication, teamwork, and problem-solving. It will be beneficial to incorporate real-world and problem solving skills into the curriculum such as using project-based learning \cite{kokotsaki2016project} where student work in teams to simulate real-world collaboration, adding communication-related assignments that involves presenting technical work to non-technical audiences, and inviting industry guest speakers to discuss expectations, trends, and career paths to hep student gain perspective beyond the classroom.
    \item Improved industry collaboration. There is a need to include industrial experience as part of the curriculum. This may be in form of mandatory internships  and partnering with industries to offer guaranteed placements, collaborating with the industry to develop capstone projects that mirror industrial experience, offering credit form participating in hackathons or coding challenges that simulate the intensity and hands-on nature of internships.
    \item Mentorship and peer support networks. The data shows that interns often look for advice from the peers. It may be necessary for academic departments to develop mentorship programs where alumni or senior students can mentor younger students. This may include peer-to-peer sessions, alumni engagement, or informal workshops where students share internship experiences, challenges and success.
\end{enumerate}
We believe the incorporation of these guides will prepare students better to meet the demands of the industry. The key themes suggest that enhancing career preparation, integrating soft skills, offering hands-on experience, and fostering mentorship and networking opportunities can help mitigate the challenges interns face when the go into the industry. These guidelines will help students feel more prepared and competitive when applying for internships and jobs in the industry.
\section{Threats to Validity}
This section discusses the threats that may affect the validity of our research. Based on the threats categories defined by Parker \cite{parker1993threats}, we discuss two major threats: internal threats and external threats to the validity of our work.

\subsection{Internal Threats}
The first threat is that we used posts that contain only the keywords "intern" or "internship". This will eliminate posts related to internship but did not specifically mention the keywords. Unfortunately, the PushAPI dump dataset only filters using a set of keywords. Adding other keywords will significantly increase the noise in the dataset and makes it more challenging to identify computing-related post. We believe the diversity and large number of posts used in this study is representative of the computing-related posts and can make up for the data loss.

Secondly, we used a large language model to classify the initial posts into computing and non-computing. This large language model has an accuracy of 65\%. This means that some posts may have been incorrectly labelled as computing. We then analyze the top 200 out of 19,633 subreddits, to filter out posts belonging to subreddits that are not likely to contain computing related posts. Hence, it is possible that some posts related to computing internship might have been excluded from the analysis. Furthermore, we consider only posts written with texts and excluded posts containing images despite the fact that a significant number of posts may be embedded inside an image. This is due to the significant resources required to accurately extract texts from images. Therefore, it is possible that some important conclusions may be inaccurate e.g., number of posts in a subreddit if the subreddit has a significant number of images. However, we believe this approach is necessary to ensure that a high quality dataset was used for the study.

Finally, it is possible that some posts may be assigned incorrectly to a topic during the LDA topic modeling process. However, there is no perfect solution in topic modeling that guarantees accurate classification for all documents. The manual verification process indicates that the Reddit mining approach used in this paper produced promising results. Furthermore, the overall topic can still be inferred based on the majority of posts that were correctly classified even if some of of the post were classified incorrectly.

\subsection{External Threats}
External validity refers to the applicability of out research conclusions beyond the settings of this study. The main external threat is that we used only one source (i.e., Reddit) for the data used in the study. Other online discussion forums, such as StackOverflow\footnote{\url{https://stackoverflow.com/}} could also be used to provide insights to intern discussions. This threat may affect the generalization of our findings. For example, StackOverflow is more programming-oriented and it is possible that the topics embedded in StackOverflow posts may be more technical. To address this threat, we plan to include data from StackOverflow in our future research work.

\section{Conclusion and Recommendation}
This paper has analyzed the discussions related to computing interns on the Reddit social media platform. We extracted internship related data from June 2009 to December 2023. We then used a large language model to identify computing related posts and also used subreddits information to filter out posts that are likely to have been falsely classified. The final data used for the analysis contains 143,912 Reddit posts. The final data also shows a continuous increase in the number of computing-internship related posts over the years.

We used topic modelling and sentiment analysis to analyse the data. The results shows that computing interns generally have a positive internship experience and they often post in five main topics related to academics, professional career, entertainment, social interaction, and the (software engineering) experience  during internship. This shows that the interns consider the internship experience to be important for their academics and professional career. Furthermore, the interns still have to deal with other life activities such as social interactions during an internship. Finally, the results tends to indicate that majority of the interns work as a software developer during the internship period.

In the future, we plan to incorporate datasets from more social media platforms e.g., StackOverflow and compare the analysis with the results presented in this paper to determine if the results will still be valid on non-Reddit data. We also plan to analyze the responses (comments) to the internship post and investigate if the feedback on intern posts generally have a positive sentiment like the posts themselves. Finally, we plan to investigate if there is a significant difference between the internship experience for computing interns and the internship experience for non-computing interns.

\bibliographystyle{elsarticle-num} 
\bibliography{ref}

\end{document}